\documentclass[prb,showpacs,twocolumn,floatfix,superscriptaddress]{revtex4}
\usepackage{times,amsmath,amsfonts,amssymb,latexsym}
\usepackage{graphicx,epsf}

\newcommand{\ket}[1]{| #1 \rangle}

\newcommand\Nd{\mathcal{N}}

\newcommand{\be}{\begin{equation}}
\newcommand{\ee}{\end{equation}}
\newcommand{\ba}{\begin{eqnarray}}
\newcommand{\ea}{\end{eqnarray}}
\newcommand{\beq}{\begin{equation}}
\newcommand{\eeq}{\end{equation}}
\newcommand{\bea}{\begin{eqnarray}}
\newcommand{\eea}{\end{eqnarray}}

\newcommand{\ignore}[1]{}
\def\CC{{\rm\kern.24em \vrule width.04em height1.46ex depth-.07ex
    \kern-.30em C}}
\def\P{{\rm I\kern-.25em P}}
\def\RR{{\rm
         \vrule width.04em height1.58ex depth-.0ex
         \kern-.04em R}}
\def\bbbone{{\mathchoice {\rm 1\mskip-4mu l} {\rm 1\mskip-4mu l}
{\rm 1\mskip-4.5mu l} {\rm 1\mskip-5mu l}}}
\def\bbbc{{\mathchoice {\setbox0=\hbox{$\displaystyle\rm C$}\hbox{\hbox
to0pt{\kern0.4\wd0\vrule height0.9\ht0\hss}\box0}}
{\setbox0=\hbox{$\textstyle\rm C$}\hbox{\hbox
to0pt{\kern0.4\wd0\vrule height0.9\ht0\hss}\box0}}
{\setbox0=\hbox{$\scriptstyle\rm C$}\hbox{\hbox
to0pt{\kern0.4\wd0\vrule height0.9\ht0\hss}\box0}}
{\setbox0=\hbox{$\scriptscriptstyle\rm C$}\hbox{\hbox
to0pt{\kern0.4\wd0\vrule height0.9\ht0\hss}\box0}}}}
\def\bbbz{{\mathchoice {\hbox{$\sf\textstyle Z\kern-0.4em Z$}}
{\hbox{$\sf\textstyle Z\kern-0.4em Z$}}
{\hbox{$\sf\scriptstyle Z\kern-0.3em Z$}}
{\hbox{$\sf\scriptscriptstyle Z\kern-0.2em Z$}}}}

\newcommand{\xx}{\mathbf x}
\newcommand{\yy}{\mathbf y}
\newcommand{\kk}{\mathbf k}
\newcommand{\rr}{\mathbf r}
\def\ellp{{\ell^{\prime}}}

\begin{document}

\title{The toric-boson model: Toward a 
topological quantum memory at finite temperature}

\author{Alioscia Hamma}
\affiliation{Perimeter Institute for Theoretical Physics, 
31 Caroline Street North, Waterloo, Ontario, Canada N2L 2Y5}
\affiliation{Research Laboratory of Electronics, Massachusetts Institute of Technology, 
 77 Massachusetts Avenue, Cambridge, Massachusetts 02139, USA}

\author{Claudio Castelnovo}
\affiliation{Rudolf Peierls Centre for Theoretical Physics, 
University of Oxford, 1 Keble Road, Oxford OX1 3NP, United Kingdom}

\author{Claudio Chamon}
\affiliation{Department of Physics, Boston University, 590 Commonwealth Avenue, 
Boston, Massachusetts 02215, USA}

\begin{abstract}
We discuss the existence of stable topological quantum memory at finite 
temperature. At stake here is the fundamental question of whether it is in 
principle possible to store quantum information for macroscopic times without the
intervention from the external world, that is, without error correction. We
study the toric code in two dimensions with an additional bosonic field that
couples to the defects, in the presence of a generic environment at finite
temperature: the \emph{toric-boson model}. 
Although the coupling constants for the bare model are not finite in the 
thermodynamic limit, the model has a finite spectrum. 
We show that, in the
topological phase, there is a finite temperature below which
open strings are confined and therefore the lifetime of the memory can
be made arbitrarily (polynomially) long in system size. 
The interaction with the bosonic field yields a long range attractive force 
between the end points of open strings, but leaves closed strings and 
topological order intact. 
\end{abstract}

\maketitle 
%
%

\section{Introduction}
In recent years, it has become clear that 
models of computation based on quantum mechanics are radically different 
from classical ones.~\cite{qcbook,qip}
However, the quantum wave function of a macroscopic system
rapidly decoheres because of the interaction with the
environment.~\cite{decoherence} Decoherence is responsible for the
appearance of a classical world.~\cite{zurek,joos} Therefore, if we
want to process quantum information, we must preserve it from being
spoiled by decoherence. 

In the past $15$ years, a great amount of literature in quantum 
information has been devoted to the theory of error correction, using several 
different paradigms.~\cite{qcbook, ec,qip}
We encounter here a profound difference with respect to the classical case: 
encoding classical information in a system can be robust, without the need 
for error correction (e.g., magnetic hard drives). 
On the other hand, the fragility of quantum states to decoherence seems to require quantum 
error correction. 
Is that really so? Can one build robust quantum memory at the physical level, 
just like one does with classical memory, or does quantum mechanics forbid it? 

For suitably symmetric models of the environment, there exist 
\textit{decoherence-free subspaces} (DFSs).~\cite{dfs}
However, it is controversial whether these subspaces can exist in practice, 
because the conditions to obtain DFSs are unstable.~\cite{nodfs}
Moreover, DFSs only exist for a specific model of noise, while we are 
interested in the question of principle whether a quantum system can be 
protected from a generic interaction with the environment, without operating 
from the external world (i.e., error correction). The only restriction we shall assume about the way the environment interacts with the system is the \emph{locality} of the interaction. 

If the only requirement is locality of
the interaction between the environment and the system, we can imagine
encoding information in some collective, \textit{topological} degrees of
freedom, and obtaining the desired robustness at the
physical level. Physical systems that exhibit topological
order seem to have the appropriate topological features~\cite{wentop}. 
The toric code is an example of a {\em self-correcting} quantum memory, 
that is, a quantum memory that is inherently robust against arbitrary
(local) perturbations without need for error correction 
\emph{at zero temperature}.~\cite{kitaev, dennis} 

However, robustness against perturbations is not enough.
What one needs is a system in which quantum information can be stored for 
macroscopically long times, in the presence of arbitrary local interactions with 
the environment at finite temperature $T$. There are not only errors due 
to virtual processes but, even more importantly, errors due to thermal 
excitations. 
For example, the toric code in two and three dimensions ($2D$s and $3D$s) is not 
stable.~\cite{dennis,Nussinov2006,castelnovo,alicki1}
[See Ref.~\onlinecite{no-go} for a discussion on generic stabilizer codes.] 
The toric code in four dimensions is stable,~\cite{dennis,castelnovo,alicki4} 
but interactions in four spatial dimensions are hardly realistic.

One wonders whether quantum information
is not only logically different from the classical one, but also
\textit{physically} different in that it cannot be stored efficiently
in a passive way 
(in less than four dimensions). 
Therefore, we are faced with a fundamental question: does a stable quantum 
memory exist at all or the laws of nature forbid it in three or less 
dimension? 

In this paper, we take a step toward realizing topological quantum memories 
in \emph{two spatial dimensions} that are stable against decoherence towards 
an arbitrary environment. We obtain this result by adding to the toric code a 
local interaction between the defects and a bosonic field 
({\em toric-boson} model). 
This result comes at the cost of a compensating defect energy 
whose strength diverges with growing system size. 
However, the protection arising in the toric-boson model is not trivially 
equivalent to removing thermal defects by introducing an 
infinite energy cost by itself. 
In our model, the physical energy scales (namely, the excitation spectrum) 
remain {\em finite}, and it is the induced long-range interactions between 
defects that render the quantum state stable to thermal fluctuations. 
This is similar to what one usually does in high-energy physics, where 
infinite coupling constants are used (often needed) to obtain finite energy 
spectra. 

We show that the toric-boson model realizes a quantum memory that can be 
stored for macroscopically long times, in the sense that the relaxation times 
scale (polynomially) with system size. 
While we fall short of realizing a stable topological quantum memory using 
\emph{local finite} coupling terms in the Hamiltonian, our model shows 
that quantum relaxation times scaling with system size can be obtained in 
a system  with a finite excitation spectrum, using 
\emph{local coupling terms only}.

%
%


\section{Toric code}

The toric code model \cite{kitaev} is defined on a $2D$ $L\times L$ square 
lattice with periodic boundary conditions, and with spin-$1/2$ degrees of freedom 
on the $N=2L^2$ edges. On every vertex 
$s$ we can define the star operator $X_s=\prod_{j\in s} \sigma^x_j$ and on every 
plaquette $p$ the plaquette operator $Z_p =\prod_{j\in p} \sigma^z_j$. The Hamiltonian 
of the toric code is
\be
H_{\rm toric} = -\frac{\Delta}{2}\left( \sum_s X_s + \sum_p Z_p \right) 
. 
\label{eq: toric}
\ee
The operators $X_s,Z_p$ all commute and therefore their eigenvalues are good 
quantum numbers for labeling the states. 
Nevertheless they are not all independent because periodic boundary conditions 
imply $\prod_s X_s=\prod_p Z_p =\bbbone$. The two degrees of freedom not 
labeled by the above quantum numbers are the two topological qubits defined 
by the algebra of the Pauli operators 
$
\{
X_{L_1} = \prod_{j\in L_1} \sigma_j^x, 
Z_{M_2} = \prod_{j\in M_2} \sigma_j^z,
\}
$ 
and $\{X_{L_2}, Z_{M_1}\}$ defined similarly, where $L_1, L_2$ and $M_1,M_2$ 
are appropriately defined non-contractible loops around the torus. 

Two topological qubits can then be defined as the subspaces on which the two 
Pauli algebras $(Z_{M_1},\,X_{L_2})$ and $(Z_{M_2},\,X_{L_1})$ act, 
respectively. 
For example, one can choose the two basis states for the first qubit to be 
the eigenvectors $\ket{0}$ and $\ket{1}$ of $Z_{M_1}$, with eigenvalues 
$+1$ and $-1$. 
In this case, the operator $X_{L_2}$ flips the qubit 
($X_{L_2}\,\ket{0}=\ket{1}$), while the operator $Z_{M_2}$  can be used to 
change the phase of a qubit state 
[e.g., $Z_{M_2}(\ket{0}+\ket{1})/\sqrt{2}=(\ket{0}-\ket{1})/\sqrt{2}$].

Since the model has a spectral gap $\Delta$, the information encoded in the 
topological qubits is stable against arbitrary local 
perturbations.~\cite{kitaev}
However, it is fragile when interacting with an arbitrary environment at finite temperature
(even if it is local).~\cite{dennis,Nussinov2006,castelnovo,alicki1} 
A heuristic argument for the fragility can be formulated as follows. 
Consider for concreteness
$
H_{\rm I} 
= 
g \sum_{j=1}^N (\sigma^x_j \otimes f_j +  \sigma^z_j \otimes f_j)
$ 
to be the interaction Hamiltonian between the system and the environment. 
This interaction will create pairs of defects (anyons) with 
probability $P\sim\exp(-\Delta/k_B T)$ and move them about in a Brownian 
motion. When two anyons travel around the non-contractible holes of the torus 
to recombine and annihilate, a logical error is produced, namely a 
non-contractible loop. 
[For example, a pair of $Z_p$ defects (generated by $\sigma^x$ operators) that 
travel around a loop of type $L_2$ will change the eigenvalue of $Z_{M_1}$, and
$\forall M_1$.] 
Since the anyons propagate freely, the time scale for an 
isolated pair to induce an error is proportional to $L^2$. 
However, the very same non-interacting nature that ensures free motion of 
the defects causes a non-vanishing density of them at any given 
temperature. 
Therefore, one can envisage faster collective processes whereby pairs 
stretch only as far as reaching anyons from neighboring pairs, and by 
annihilating them, they lead to longer overall propagation paths. 
The time scale to induce an error via these processes is roughly given by the 
time it takes for a pair to stretch over a distance equal to the average 
inter-particle separation, $t_{\rm rel}\sim\exp(\Delta/k_B T)$. 
The relaxation time is therefore microscopic, because it does not scale 
with system size.

The reason why end-points of strings (the anyons) can propagate freely is 
that here only the boundary of string-like objects pays an energy, and a longer 
string does not have a larger boundary. So one way of confining these objects 
is to go to higher dimensions. 
In the 3D toric code, topological information is stored in non-contractible 
membranes and winding loops. Defects to these operators acquire the form of 
open strings and open membranes, respectively. 
While open strings still behave as in 2D, with end-points propagating 
freely, an open membrane costs energy that grows with the size of its boundary
(a closed string), 
and it is therefore confined. 
Since only one of the two operators is preserved at finite temperature, 
the 3D toric code can only encode classical 
information~\cite{castelnovo,alicki1}. 
In 4D, true quantum memory can be stored for times that are exponentially 
large in $L$, because both types of errors are 
membranes.~\cite{dennis,castelnovo,alicki4}

One can think of confining the strings by adding a tension term to the 
Hamiltonian. There is a critical value of the coupling with this field such 
that the open strings are confined~\cite{num}. 
Unfortunately, the tension term does not distinguish 
between open and closed strings. If open strings are confined, the system 
undergoes a quantum phase transition where topological order is destroyed, and 
with it the degenerate ground state encoding the qubit is also lost. 
In fact, the phase in which the strings are confined can be mapped to the 
ferromagnetic phase of the $2D$ Ising model~\cite{HammaLidar:06}, and we can 
only encode classical information in such system.
%
%

\section{Toric-boson model
        }
If we want to hinder the propagation of the anyons without affecting the loops 
and spoiling topological order, we must find an interaction that only couples 
the end-points of strings. A projector onto the subspace of open strings would 
allow us to give a tension only to open strings. But this term would be highly 
non-local, and therefore not realistic. Another possibility is to add a 
long-range force that makes the anyons attract, as it was suggested in 
Ref.~\onlinecite{dennis}. However, the problem is whether the needed attractive 
force can materialize from a purely local Hamiltonian.

As we show in the following, one can reach this goal with a local Hamiltonian 
by coupling the 2D toric code to bosonic fields, which will have the desired 
confining effect in the presence of a completely generic bath acting on both 
systems. 
However, we shall achieve this at the cost of a compensating energy term 
in the Hamiltonian that scales with the size of the system. The bosonic fields will affect open strings by attracting the end-points, 
while they have no effect whatsoever on closed strings. In this way, they 
protect the topological qubit while preserving topological order. 

Let $n_s$ and $n_p$ be the defect density operators [namely, $n_s=(1-X_s)/2$ 
and $n_p=(1-Z_p)/2$, respectively] taking values $0$ and $1$ on the stars / 
plaquettes of a square lattice (with spacing set to unity). 
This density will couple locally to the displacement field defined in the 
continuum. We consider the case of elastic waves (phonons) on 
the surface of the torus, two in plane and one out of plane. 
These elastic deformations are denoted by 
$\varphi(\xx,t),\vec{\phi}(\xx,t)=(\phi^x(\xx,t),\phi^y(\xx,t))$, where 
$\xx$ takes values in the continuum [with centers
of plaquettes labeled by $\xx_p$, and those of stars by $\xx_s$]. 

The densities $n_s$ and $n_p$ couple to the out-of-plane phonons 
piezoelectrically ($\varphi$)~\cite{Mahan}, 
and to the in-plane phonons in a rotationally 
invariant form (via $\partial_{x} \phi^x + \partial_y \phi^y$). 
In the case of piezoelectric coupling, one has to take proper care of the 
center of mass (CM) mode. One can work with 
$\varphi_{\kk=0}$ 
pinned to the origin (basically, the entire mass of the crystal behaves as a 
large classical object). This is physical, since what matters in the piezo
coupling are the elastic deformations, and uniform translations should
not affect the system; by working with coordinates relative to the CM,
this is ensured.

One can expand $\varphi(\xx,t),\phi^i(\xx,t)$ using creation and annihilation 
operators $a^{}_{\kk},a^{\dagger}_{\kk},b^{(i)}_{\kk},b^{(i)\dagger}_{\kk}$ 
satisfying the harmonic oscillator algebra as
\begin{eqnarray}\label{fields}
\varphi(\xx,t)
&=&
\frac{1}{\sqrt{V}} \sum_{\kk\ne 0} \frac{1}{\sqrt{2\omega_\kk}}\;
\left[
a^{}_{\kk}(t)\: e^{i\kk\cdot\xx}
+
a^{\dagger}_{\kk}(t)\: e^{-i\kk\cdot\xx}
\right]
\nonumber \\ 
\label{eq: varphi field}
\end{eqnarray}
and
\begin{eqnarray}
\phi^i(\xx,t)
&=&
\frac{1}{\sqrt{V}} \sum_{\kk\ne 0} \frac{1}{\sqrt{2\Omega_\kk}}\;
\left[
b^{(i)}_{\kk}(t)\: e^{i\kk\cdot\xx}
+
b^{(i)\dagger}_{\kk}(t)\: e^{-i\kk\cdot\xx}
\right] 
,
\nonumber \\ 
\label{eq: phi fields}
\end{eqnarray}
with $i=x,y$, thus implying the canonical equal-time commutation relations
\begin{eqnarray}
\nonumber
[\phi^i(\xx,t),\dot\phi^i(\xx',t)]&=&i\delta(\xx-\xx'),  \\
\left[\varphi(\xx,t),\dot\varphi(\xx',t)\right]&=&i\delta(\xx-\xx')
\end{eqnarray}
for the displacement fields.
The Hamiltonian for the fields is given by 
\begin{eqnarray}
H_{\rm boson} 
&=& 
\sum_{\kk\ne 0} \omega_\kk\;a^{\dagger}_{\kk} a^{}_{\kk}
+
\sum_{\kk\ne 0,i=x,y} \Omega_\kk\;b^{(i)\dagger}_{\kk} b^{(i)}_{\kk} .
\label{hboson}
\end{eqnarray}
The interaction Hamiltonian reads as
\begin{equation}
H_{\rm int} 
= 
\sum_{\ell=s,\,p} n_\ell 
\left\{\vphantom{\sum}
  g_\omega \, \varphi(\xx_\ell)
  +
  g_\Omega \left[ \partial_x\phi^x(\xx_\ell)+\partial_y\phi^y(\xx_\ell) \right]
\right\}
\label{hint}
\end{equation}
(notice that $g_\omega$ and $g_\Omega$ have different dimensions), 
and the toric-boson Hamiltonian can thus be written as
\be
H = H_{\rm toric}+H_{\rm boson}+H_{\rm int} 
. 
\label{hmodel}
\ee

After substituting Eqs.~(\ref{eq: varphi field}) and~(\ref{eq: phi fields}) 
into the interaction Hamiltonian~(\ref{hint}), and introducing new creation 
and annihilation operators 
\bea
\alpha^{\ }_{\kk}
&\equiv&
a^{\ }_{\kk}
+
\frac{g_\omega \, \tilde\rho^\dagger(\kk)}{\sqrt{2V}\,{\omega_\kk}^{3/2}}
\\ 
\beta^{(i)}_{\kk}
&\equiv&
b^{(i)}_{\kk}
+
\frac{i g_\Omega \kk_i \, \tilde\rho^\dagger(\kk)}{\sqrt{2V}\,{\Omega_\kk}^{3/2}}
\qquad 
(i = x,y)
\:, 
\eea
the total Hamiltonian $H$ of the system can be written as 
\begin{eqnarray}
H
&=&
H_{\rm toric} 
+ 
\sum_{\kk\ne 0} \omega_\kk\;\alpha^{\dagger}_{\kk} \alpha^{}_{\kk}
+
\sum_{\kk\ne 0,i=x,y} \Omega_\kk\;\beta^{(i)\dagger}_{\kk} \beta^{(i)}_{\kk} 
\nonumber \\
&-& 
\frac{1}{V}\sum_{\kk\ne 0} 
  \left\vert \tilde\rho(\kk) \right\vert^2 
  \left(
    \frac{g^2_\omega}{2\,{\omega_\kk}^2}
    +
    \frac{g^2_\Omega (k^2_x + k^2_y)}{2\Omega_\kk^2}
  \right)
\:,
\label{htotal}
\end{eqnarray}
where $\tilde\rho(\kk) = \sum_\ell n_\ell e^{i \kk\cdot\xx_\ell}$ is the 
Fourier transform of $\rho(x) = \sum_\ell n_\ell \, \delta(\xx - \xx_\ell)$. 
Here we have assumed that the defects are smeared over the corresponding 
plaquette / star, so as to regularize $\delta(\xx)$ and remove ultraviolet 
divergences.

Notice that the first two terms in Eq.~(\ref{htotal}) lead to
precisely the same energy levels as those of the bosonic modes in the
absence of the defects, [Eq.~(\ref{hboson})]. 
This holds true even for thermal averages in mixed states. 
Therefore, the phononic fields added to the toric code Hamiltonian give rise 
to an effective interaction potential $V_d$ given by the last term in 
Eq.~(\ref{htotal}), irrespective of temperature. 
Notice that $V_d$ is always negative, independent of the arrangement of 
the defects, and thus it favors defect proliferation, which will 
need to be appropriately counter-balanced later on. 

Consider the case of $\Nd\equiv \Nd_X+\Nd_Z$ defect pairs, at positions 
$\xx_a$, with $\ell=1,\dots,2\Nd_X$ labeling the sites with star-type defects 
and $\ell=2\Nd_X+1,\dots,2\Nd$ labeling the sites with plaquette-type 
defects. 
The effective interaction potential $V_d$ can then be written as 
\begin{eqnarray}
 V_d\left(\xx_1,\dots,\xx_{2\Nd}\right) &=& - \sum_{\ell,\ellp=1}^{2\Nd} n_\ell n_{\ellp} \frac{1}{V}\sum_{\kk\ne 0} \\
&&\times \left[
  \frac{g^2_\omega}{2\,\omega^2_\kk}
  +
  \frac{g^2_\Omega \vert\kk\vert^2}{2\,\Omega^2_\kk}
\right]
e^{i\kk\cdot(\xx_\ell-\xx_\ellp)}
\quad\;\;
\nonumber
\\
&=&
\frac{1}{2}
\sum_{\ell,\ellp=1}^{2\Nd} n_\ell n_\ellp 
\left[\vphantom{\sum}
  V_\omega(\xx_\ell-\xx_\ellp) + V_\Omega(\xx_\ell-\xx_\ellp)
\right], 
\end{eqnarray}
where 
\bea
V_\omega(\xx_\ell-\xx_\ellp) 
&\equiv& 
-
\frac{1}{V} \sum_{\kk\ne 0} 
  \frac{g^2_\omega}{\omega^2_\kk} e^{i\kk\cdot(\xx_\ell-\xx_\ellp)},
\label{eq: V_omega}
\\ 
V_\Omega(\xx_\ell-\xx_\ellp) 
&\equiv& 
-
\frac{1}{V} \sum_{\kk\ne 0} 
  \frac{g^2_\Omega \vert\kk\vert^2}{\Omega^2_\kk} e^{i\kk\cdot(\xx_\ell-\xx_\ellp)}.
\label{eq: V_Omega}
\eea
For acoustic phonons with dispersion $\omega_\kk=v_\omega \vert\kk\vert$, 
and $\Omega_\kk=v_\Omega \vert\kk\vert$, the potential $V_\omega(\rr)$ becomes 
Coulombic in $d$-dimensions, 
\beq
\nabla^2\,V_\omega(\rr) 
=
\frac{g^2_\omega}{v_\omega^2} 
  \frac{1}{V} \sum_{\kk\neq0} e^{i\kk\cdot\rr} 
=
\frac{g^2_\omega}{v_\omega^2} \left[ \delta^{(d)}(\rr) - \frac{1}{V} \right] 
. 
\eeq
Namely, the potential generated by a point-like charge in a uniform 
compensating background distribution is
\bea
V_\omega(\rr) 
&=& 
\begin{cases}
\frac{g^2_\omega}{v_\omega^2} 
  \frac{\Gamma\left({d}/{2}\right)}{2 \pi^{d/2}} \frac{1}{2-d} \, 
    \vert\rr\vert^{2-d} + C
& \quad\text{$d\ne 2$}
\\
& \\
\frac{g^2_\omega}{v_\omega^2} \frac{1}{2 \pi} \ln \vert\rr\vert + C
& \quad\text{$d= 2$} 
,
\end{cases}
\nonumber 
\eea
where the uniform compensating background is irrelevant 
due to the absence of the $\kk=0$ term in Eq.~(\ref{eq: V_omega}).
For the same reason, $\int\!V_\omega(\xx)\,d^dx = 0$, which fixes the 
constant $C \equiv - g^2_\omega / (2 \pi v_\omega^2) \ln \xi_L$, 
with $\xi_L$ of the order of the system size 
$L$ in two dimensions (the actual value will depend on details about the 
shape of the system). 

For $g_\omega > 0$, the potential $V_\omega(\rr)$ acts therefore as 
as a ``gravitational'' potential with Newton constant 
$G=(2v_\omega^2 \pi^{d/2})^{-1} \Gamma\left({d}/{2}\right)$, attracting 
particles of mass $g_\omega$. 

The in plane phonon contribution amounts to 
\beq
V_\Omega(\rr) 
= 
-
\frac{g^2_\Omega}{v_\Omega^2} 
  \frac{1}{V} \sum_{\kk\neq0} e^{i\kk\cdot\rr} 
=
- \frac{g^2_\Omega}{v_\Omega^2} \left[ \delta^{(d)}(\rr) - \frac{1}{V} \right] 
, 
\eeq
and we can finally combine the two contributions into 
\begin{subequations}
\bea
V_\Omega(\rr) + V_\omega(\rr) 
&=& 
\frac{g^2_\omega}{v_\omega^2} \frac{1}{2 \pi} \ln \frac{\vert\rr\vert}{a} 
\label{eq: V_Omega + V_omega 1} 
\\ 
&+& 
\left[ 
  \frac{g^2_\Omega}{v_\Omega^2} \frac{1}{V} 
  - 
  \frac{g^2_\omega}{v_\omega^2} \frac{1}{2 \pi} \ln \frac{\xi_L}{a} 
\right] 
\label{eq: V_Omega + V_omega 2} 
\\ 
&-& 
\frac{g^2_\Omega}{v_\Omega^2} \delta^{(2)}(\rr) 
, 
\label{eq: V_Omega + V_omega 3} 
\eea
\end{subequations} 
where we separated for convenience $\ln (\vert\rr\vert / \xi_L)$ into 
$\ln (\vert\rr\vert / a) - \ln (\xi_L / a)$ 
($a$ being the lattice constant, $\vert\rr\vert / a > 1$, $\forall\,\rr$). 
The first term is the desired confining interaction (which we discuss 
below), while terms~(\ref{eq: V_Omega + V_omega 2}) 
and~(\ref{eq: V_Omega + V_omega 3}) are by-products that need to be 
appropriately taken care of. 
By tuning the ratio between $g_\omega/v_\omega$ and $g_\Omega/v_\Omega$ one 
can obtain that the two parts of Eq. (\ref{eq: V_Omega + V_omega 2}) 
cancel each other out exactly, at the cost however of generating a system-size 
scaling coupling constant in Eq. (\ref{eq: V_Omega + V_omega 3}): 
\beq
\frac{g^2_\Omega}{v_\Omega^2} 
= 
\frac{g^2_\omega}{v_\omega^2} \frac{V}{2 \pi} \ln \frac{\xi_L}{a} 
\sim 
L^2 \ln L 
. 
\eeq
Therefore, the phonon-mediated effective interaction between the defects 
can be written as 
\begin{eqnarray}
&& V_d\left(\xx_1,\dots,\xx_{2\Nd}\right) = 
\nonumber \\ 
&& \quad 
=
\frac{g^2_\omega}{v_\omega^2} \frac{1}{4 \pi} 
  \sum_{\ell,\ellp=1}^{2\Nd} n_\ell n_\ellp 
    \ln \frac{\vert \xx_\ell-\xx_\ellp \vert}{a} 
\nonumber \\ 
&& \quad 
- 
\frac{g^2_\omega}{v_\omega^2} \frac{1}{4 \pi} 
\left( V \ln \frac{\xi_L}{a} \right) 
  \sum_{\ell,\ellp=1}^{2\Nd} n_\ell n_\ellp \: 
    \delta^{(2)}(\xx_\ell-\xx_\ellp) 
\\ 
&& \quad 
=
\frac{g^2_\omega}{v_\omega^2} \frac{1}{4 \pi} 
  \sum_{\ell,\ellp=1}^{2\Nd} n_\ell n_\ellp 
    \ln \frac{\vert \xx_\ell-\xx_\ellp \vert}{a} 
\nonumber \\ 
&& \quad 
- 
\frac{g^2_\omega}{v_\omega^2} \frac{1}{4 \pi} 
\left( V \ln \frac{\xi_L}{a} \right) 
  2 \Nd \: \delta^{(2)}(0) 
, \quad 
\label{eq: effective pot + chem pot}
\end{eqnarray}
where the $\delta$-function has been regularized at short distances
just like one does for the Coulomb potential. 

The remaining unwanted term -- 
coming from Eq. (\ref{eq: V_Omega + V_omega 3}) -- 
plays the role of a chemical potential that favors thermal defects, and 
whose strength scales with the size of the system. 
In order to preserve the fact that the defect density vanishes at zero 
temperature, we must neutralize this contribution by adding a compensating 
energy cost for the defects that also grows with system size. 

Since this unwanted term scales linearly in the number of defects $\Nd$, 
it can be removed by the further addition of a chemical potential to the 
Hamiltonian of the system. 
Indeed, the role of the in plane phonons in our model is to prevent the arising  
of undesired defect-favoring long-range interactions that scale with 
the square of the number of defects $\Nd^2$, and could not be removed by the 
addition of a local one-body term in the Hamiltonian. 

In conclusion, we obtain a purely gravitational defect interaction of the 
form 
\begin{eqnarray}
V_d\left(\xx_1,\dots,\xx_{2\Nd}\right) 
=
\frac{g^2_\omega}{v_\omega^2} \frac{1}{4 \pi} 
  \sum_{\ell,\ellp=1}^{2\Nd} n_\ell n_\ellp 
    \ln \frac{\vert \xx_\ell-\xx_\ellp \vert}{a} 
. 
\end{eqnarray}

An important comment is in order. 
While we did add an infinite ($\sim L^2 \ln L$) coupling term to the 
Hamiltonian in order to arrive at the effective interaction shown above, 
this is not equivalent to trivially introducing an infinite energy cost for the 
defects, which would forbid their presence altogether. 
In fact, the energy to add a pair of defects in our model remains 
\emph{finite} 
(the environment acts locally on the system and therefore defects are always 
produced close to each other, and move one step at a time). 

The situation is similar in high energy, where the bare Hamiltonian often 
has diverging coupling constants, but it carries nonetheless physical meaning 
as the spectral gaps are finite. 

%
%

\subsection{
Robustness of quantum memory
           }
Once we obtain an effective long-range ``gravitational'' attraction between 
the defects, we can argue for the stability of the quantum memory. 
It is well known that the partition function of $\Nd$ particles of mass $m$ 
interacting via a 2D gravitational potential $u(r) = G m^2 \ln(r)$ becomes 
divergent below a finite temperature 
$T_* = \Nd G m^2 / 4$ (see e.g., Ref.~\onlinecite{grav_pot}). 
The gravitational forces overcome the entropic contribution and lead to a 
collapse, where all the particles coalesce to a single point. 
In the toric-boson model, where the massive particles correspond to the 
defective stars and plaquettes, their number is controlled by the chemical 
potential due to the ratio $\Delta/T$. 
Therefore, given that the number of particles is topologically constrained to 
be even, the equilibrium density of defects vanishes below the finite 
temperature $T_*(\Nd=2) = G m^2 / 2$. 

In this temperature regime, thermal processes connecting two different 
topological sectors are essentially limited to creation, diffusion and 
annihilation of individual pairs of particles. Such processes must overcome 
the energy barrier of taking one particle around the whole system, which 
scales with the logarithm of the system size, thus yielding a characteristic 
\emph{macroscopic} time scale 
$t_{\rm rel}\sim\exp[\alpha \ln(L)/T] \sim L^{\alpha/T}$ which is polynomial 
in system size $L$ [$\mbox{poly}(L)$]. 
%
%

\section{
Robustness of the toric-boson model
        } 
An essential ingredient to obtain the effective gravitational 
potential between the defects lies in the gaplessness of the phonon modes 
involved in the system. 
In order to ensure robustness against thermal fluctuations in the toric-boson 
model, one should therefore also ensure that the action of the environment on the 
phonon modes does not open a gap (e.g., via introducing effective 
interactions between the different bosonic modes). 

In our case, the gaplessness ($\omega_\kk \sim |\kk|$) of phononic 
excitations is protected by translational symmetry. 
Even if the elastic medium is inhomogeneous, it satisfies the equation
of motion $m(\xx)\,\ddot\phi(\xx,t)=\int d^dy \:K(\xx,\yy)\,\phi(\yy,t)$, 
where the condition $\int d^dy \:K(\xx,\yy)=0$ 
follows from uniform translations $\phi(\xx,t)={\rm const.}$; 
this condition on the $K(\xx,\yy)$ is similar to that in a
master equation for stochastic systems~\cite{OrbachRMP}, with the
difference that the wave equation is second order. It thus follows
from generic arguments on diffusion in inhomogeneous media that the
dispersion of the phonons for long wavelengths must be linear,
$\omega_\kk \sim |\kk|$. 
Notice that this conclusion holds even if the inhomogeneous
elastic medium breaks translation symmetry [in the form of an
inhomogeneous $m(\xx)$ and $K(\xx,\yy)$]. The key point is that, at
long distances, acoustic sound waves propagate in the effective
medium. Notice also that to attempt to pin the displacement fields with a
rigid substrate potential is an illusion, as one must consider the
elastic properties of the substrate itself. The linearly dispersing
acoustic modes in the effective medium at large distances are
responsible for the the long-range behavior of $v(\rr)$, which
remains unchanged after tracing out the environment. 
[One can bypass the issue of the linear dispersion and compute
directly $V_\omega(\rr)$ in the inhomogeneous medium:
$V_\Omega(\xx',\xx)=\int_0^\infty dt\,P_W(\xx',\xx;t)$, where
$P_W(\xx',\xx;t)$ is the probability to diffuse from $\xx$ to $\xx'$
in time $t$, computed using the transition rate
$W(\xx,\yy)=K(\xx,\yy)/m(\xx)$.]
%
%
%

\section{
Conclusions and outlook
        }
In this paper, we have shown how a system with topological order like the 
toric code can be made robust to local interactions with a generic 
environment below a finite temperature $T_*$, if the defects in the system are 
coupled to a phononic field. 
The defects feel a long-range attractive force that prevents them from 
moving freely around the torus. The long-range force is not spoiled by the 
interaction of the bosons with the environment as long as the dispersion 
relation is preserved. In our model, phonons are protected by 
translational symmetry and the dispersion relation yields a logarithmic 
potential resulting in a relaxation time scaling as $\mbox{poly}(L)$. 
Therefore, quantum information can be stored for arbitrarily long times 
scaling with the size of the system. 

In the model described here, we need to introduce compensating terms that 
scale as $L^2 \ln L$, $L$ being the linear size of the system, and thus 
diverge in the thermodynamic limit. 
However, we do so while preserving the finite nature of the low-energy 
spectrum of the system. 
We believe that this is a step farther with respect to the simple addition of 
an infinite energy cost for the defects in the Hamiltonian. 
Our model shows that, in principle, quantum mechanics allows for the 
preservation of quantum information for arbitrarily long times, scaling with 
the size of the system. 

There are a number of open problems left to investigate. 
First and foremost, can one construct a model with similar physics where all 
the interactions are not only local but also \emph{finite}? 

Second, for reliable quantum topological memory, $\mbox{poly}(L)$ is not ideal. 
It takes in general polynomial time to operate on a topological qubit 
(e.g., the read-out at finite temperature~\cite{alicki1}). 
Our system would be viable only in the low temperature limit, where the 
polynomial relaxation time scale $t_{\rm rel}\sim L^{\alpha/T}$ is much 
larger than the (temperature independent) polynomial operational time scale. 
A stronger quantum memory could be realized in a system that exhibits 
exponential relaxation time scales $t_{\rm rel}\sim \exp (L)$. 
This could be obtained using a bosonic field with 
protected gapless excitations with dispersion $\omega\sim |k|^3$. 
There are other 
topologically ordered systems with the right cubic dispersion relation, whose 
gaplessness is protected by topological order itself, resulting in a very 
strong protection~\cite{wenprivate}. Therefore, a toric code interacting with 
the appropriate system would feature exponentially large relaxation times. 
We think that the robustness of quantum memory in such system is the true 
meaning of the full survival of quantum topological order at finite 
temperature. 

%
%

\section{Acknowledgments} 
We acknowledge important discussions with
M.~Hastings, X.-G. Wen and D. Lidar.  We are particularly grateful to
M.~Hastings for pointing out some of the undesired terms that arise in
our Hamiltonian. Such terms would have led to phase separation effects
in the system, and we had missed them in a previous version of the
manuscript.  Research at Perimeter Institute for Theoretical Physics
is supported in part by the Government of Canada through NSERC and by
the Province of Ontario through MRI.  This work was supported in part
by EPSRC under Grant No. GR/R83712/01 (C.~C.), and by a grant from
the Foundational Questions Institute (fqxi.org), a grant from xQIT at
MIT (A.~H.). A.~H. and C.~C. thank the BU Condensed
Matter Theory visitors program for its hospitality.
%
%


\end{document}